\documentclass{webofc}\usepackage[varg]{txfonts}
\usepackage{epsfig,graphics,amssymb,amsmath,mathrsfs,color,colortbl,
bm,booktabs,cool,dcolumn}

\begin{document}\title{Bethe--Salpeter Bound-State Solutions:\\
Examining Semirelativistic Approaches}\author{Wolfgang
Lucha\inst{1}\fnsep\thanks{\email{Wolfgang.Lucha@oeaw.ac.at}}}
\institute{Institute for High Energy Physics, Austrian Academy of
Sciences, Nikolsdorfergasse 18,\\A-1050 Vienna, Austria}
\abstract{Within the formalism of relativistic quantum field
theory an adequate framework for the description of two-particle
bound states, such as, for instance, all conventional (i.e.,
non-exotic) mesons, is provided by the Poincar\'e-covariant
homogeneous Bethe--Salpeter equation. In applications, however,
this approach usually proves to be rather involved, whence it is
not always quite easy to extract the predictions sought. In view
of this, a coarse idea of the bound-state spectrum to be expected
might be gained by adhering to some simplifying approximations --
which constitutes an entirely legitimate first step. The
reliability of the insights inferred from the arising simpler
bound-state equation may be straightforwardly examined by taking
into account a couple of rigorous constraints on the obtained
discrete spectrum. Application of these tools is illustrated for
popular potentials.}\maketitle

\section{Bound States of Spinless Constituents: Semirelativistic
Approach}\label{SRA}A main issue in both relativistic quantum
physics and quantum field theories, such as quantum
electrodynamics and quantum chromodynamics, is to devise
appropriate (and, if manageable, convenient) approaches to the
bound states expected within specific settings. In this context, a
very useful tool is provided by the spinless Salpeter equation,
i.e., the eigenvalue equation of a Hamiltonian operator composed
of the relativistic kinetic energies of the involved bound-state
constituents and some static interaction potential $V$ assumed to
reflect the bound-state~relevant aspects of any underlying quantum
(field) theory. If $V$ meets these expectations, this approach
should return a rough idea of the bound-state spectrum. For two
bound-state constituents with relative coordinates $\bm{x}$ and
$\bm{p}$ in configuration and momentum space, this Hamiltonian
becomes\begin{equation}H=\sqrt{\bm{p}^2+m_1^2}+\sqrt{\bm{p}^2+m_2^2}
+V(\bm{x})\label{H}\end{equation}(where $m_1$ and $m_2$ denote
both interacting particles' masses) and, for $m_1=m_2=m$,
reduces~to\begin{equation}\widetilde H=2\,\sqrt{\bm{p}^2+m^2}+
V(\bm{x})\ .\label{H2}\end{equation}Conceptually, the spinless
Salpeter equation can be formulated along two opposite directions:
\begin{itemize}\item On the one hand, the spinless Salpeter
equation can be understood as an improvement of the
nonrelativistic Schr\"odinger equation that is achieved upon
allowing, in the Hamiltonian, the kinetic part to retain its
proper relativistic form instead of sticking to its
nonrelativistic~limit.\item On the other hand, upon starting from
the Poincar\'e-covariant \emph{Bethe--Salpeter formalism\/}
\cite{BS}, representing an appropriate framework for the
description of bound states within the realms of relativistic
quantum field theories, the spinless Salpeter equation can be
easily derived by applying several simplifying approximations to
the homogeneous Bethe--Salpeter equation. The detailed recipe
reads: discard all timelike coordinates (which leads to the
\emph{instantaneous\/} Bethe--Salpeter formalism \cite{WLe}),
require free propagation of bound-state constituents (which
produces the Salpeter equation \cite{SE}), drop all
negative-energy contributions (which results in what is called the
\emph{reduced\/} Salpeter equation) and, finally, ignore the spin
degree of freedom of all bound-state constituents in order to
eventually arrive at the \emph{spinless\/} Salpeter~equation.
\end{itemize}

\section{Analysis of Semirelativistic Hamiltonian Operators: Some
Insights}\label{SSH}The nonlocality of the Hamiltonian (\ref{H}),
induced by the square-root operators of the relativistic kinetic
energies, impedes an entirely analytic treatment of any emerging
bound-state problem. Nonetheless, by utilization of instruments
supplied by functional analysis a qualitative or even
semiquantitive picture of expectable solutions of the spinless
Salpeter equation may be drawn. In particular, establishing the
mere existence of discrete bound states by proving boundedness
from below of the spectrum of Hamiltonian operators is, beyond
doubt, of utmost importance.

\subsection{Spectrum of Semirelativistic Hamiltonian Operator:
Boundedness from Below}\label{BB}Quite generally, well-definedness
of a Hamiltonian requires, among others, this operator to be
bounded from below. Noting the manifest positivity of the
relativistic kinetic-energy operator, for semirelativistic
Hamiltonians -- such as those presented in Eqs.~(\ref{H}) or
(\ref{H2}) -- the crucial issue is the \emph{possibly singular\/}
behaviour of the interaction potentials showing up in the
Hamiltonian. In such a context, the (in some sense) worst-case
scenario originates in the Coulomb potential:\begin{equation}
V_{\rm C}(\bm{x})=V_{\rm C}(|\bm{x}|)\equiv-\frac{\alpha}
{|\bm{x}|}\ ,\qquad\alpha>0\ .\label{CP}\end{equation}For the
spinless relativistic Coulomb problem, in turn, the spectral
theory has been thoroughly worked out \cite{IWH1,IWH2}: A
semirelativistic Hamiltonian (\ref{H2}) with a Coulomb interaction
potential~(\ref{CP}) is bounded from below if and only if the
involved coupling parameter $\alpha$ satisfies the constraint
\begin{equation}\alpha<\frac{4}{\pi}=1.273239\dots\ .\label{cc}
\end{equation}Only for these strengths $\alpha$, the lower bound
to the spectrum $\sigma(\widetilde H)$ of the Hamiltonian
(\ref{H2})~reads\begin{equation}\sigma(\widetilde H)\ge2\,m\,
\sqrt{1-\left(\frac{\pi\,\alpha}{4}\right)^{\!2}}\ .\label{IWH}
\end{equation}For a more refined range, $\alpha\le1$, a somewhat
increased lower bound on $\sigma(\widetilde H)$ can be
found~\cite{MR}:\begin{equation}\sigma(\widetilde H)\ge2\,m\,
\sqrt{\frac{1+\sqrt{1-\alpha^2}}{2}}\ .\end{equation}

\subsection{Spinless-Salpeter Hamiltonian Operators: Number of
Discrete Eigenstates \cite{ICD}}\label{nde}Rigorous upper limits
on the total number of bound states described by some spinless
Salpeter equation may be formulated \cite{ICD} in terms of
(Lebesgue) spaces of measurable functions, $L^p$, on the Euclidean
space ${\mathbb R}^3$, for $p=3/2,3$. For a Hamiltonian operator
acting on the Hilbert~space $L^2({\mathbb R}^3)$ and a potential
that is a negative smooth function of compact support, hence,
subject~to\begin{equation}V(\bm{x})\in C^\infty_0({\mathbb R}^3)\
,\qquad V(\bm{x})\leq0\ ,\label{C}\end{equation}\newpage\noindent
in equal-mass two-particle problems (\ref{H2}) the total number of
bound states, $N$, is constrained~by\begin{equation}N\le\frac{C}
{48\,\pi^2}\int{\rm d}^3x\left[|V(\bm{x})|\left(|V(\bm{x})|+4\,m
\right)\right]^{3/2}\ ,\quad C=\left\{\begin{array}{rl}
14.107590867&\quad\mbox{if}\ \ \ m>0\ ,\\[.5ex]
6.074898097&\quad\mbox{if}\ \ \ m=0\ ,\end{array}\right.\label{D}
\end{equation}with numerical results for $C$ taken from
Ref.~\cite{WL:K}. Finiteness forces the potential $V(\bm{x})$
to~satisfy\begin{equation}V(\bm{x})\in L^{3/2}({\mathbb R}^3)\cap
L^3({\mathbb R}^3)\ .\label{L}\end{equation}

\subsection{Semirelativistic Hamiltonian Operators: Upper Limits on
Discrete Eigenvalues}\label{U}Upon having succeeded to establish
-- in the delicate instance of singular interaction potentials by
rather shameless abuse of the findings gained for the spinless
relativistic Coulomb problem \cite{IWH1,IWH2} (reviewed in
Subsect.~\ref{BB}) -- the rigorous boundedness from below of the
semirelativistic Hamiltonian in one's focus of interest, that is,
of that operator's spectrum, most likely one will be tempted to
try, as a next step, to acquire (at least, limited) information on
the actual location of its lowest-lying discrete eigenvalues. Such
goal can be achieved by, for instance, narrowing down the
conceivable range of any of these eigenvalues by finding an upper
bound to its range. The theoretical foundation of any attempt of
this kind is the minimum--maximum~theorem \cite{RS}:
\begin{itemize}\item Consider some self-adjoint operator, $H$,
bounded from below, with its (ordered) eigenvalues\begin{equation}
E_0\le E_1\le E_2\le\cdots\ .\label{ev}\end{equation}\item Define
the operator $\widehat H$ by \emph{restricting\/} $H$ to some
$d$-dimensional subspace of the domain~of~$H$.\item Let $\widehat
E_0,\widehat E_1,\dots,\widehat E_{d-1}$ be the $d$ eigenvalues of
the restriction $\widehat H$, likewise ordered according~to
\begin{equation}\widehat E_0\le\widehat E_1\le\widehat E_2\le\cdots
\le\widehat E_{d-1}\ ;\label{rev}\end{equation}\item these form
upper bounds to the first $d$ eigenvalues of $H$ below its
essential spectrum's~onset:\begin{equation}E_k\le\widehat E_k
\qquad\forall\quad k=0,1,2,\dots,d-1\ .\label{mini}\end{equation}
\end{itemize}

Below, in Sect.~\ref{SSP}, we will show slight predilection for
interactions governed by spherically symmetric potentials. If
dealing with central potentials, mere convenience dictates to span
the $d$-dimensional subspace required by the minimum--maximum
theorem by a basis the elements of which are functions,
$\psi_{k\ell}(\bm{x})$, that are products of a spherical harmonic
${\cal Y}_\ell$ for orbital angular momentum $\ell$ and a radial
factor involving the generalized-Laguerre polynomials
$L_k^{(\gamma)}$ \cite{WL:T1,WL:T2}:\begin{align}
\psi_{k\ell}(\bm{x})\propto|\bm{x}|^{\ell+\beta-1}
\exp(-\mu\,|\bm{x}|)\,L_k^{(2\ell+2\beta)}(2\,\mu\,|\bm{x}|)\,
{\cal Y}_\ell(\Omega_{\bm{x}})\ ,\qquad k\in{\mathbb N}_0\ ,\qquad
\ell\in{\mathbb N}_0\ ,&\label{LB}\\L_k^{(\gamma)}(x)\equiv
\sum_{t=0}^k\, \binom{k+\gamma}{k-t}\,\frac{(-x)^t}{t!}\
,\qquad\mu>0\ ,\qquad \beta>-\frac{1}{2}\ ;&\nonumber\label{LB}
\end{align}herein, $\Omega_{\bm{x}}$ indicates the $\bm{x}$-space
solid angle, and $\mu$ and $\beta$ represent two variational
parameters.

\subsection{Spinless-Salpeter Hamiltonian Operator: Quality
Assurance of its Eigenstates}Irrespective of the actual origin of
approximate eigenstates $|\chi\rangle$ of a Hamiltonian operator,
their accuracy \cite{WL:Q1,WL:Q2} can be judged (or even
quantified) by their degree of fulfilment of the \emph{master\/}
or \emph{relativistic\/} virial theorem \cite{WL:RVT1,WL:RVT2}
relating the expectation values of the radial~derivatives of all
kinetic terms and of the potential. This relation comprises, of
course, the well-known~virial theorem of nonrelativistic quantum
theory as a special case. For the operators (\ref{H}), it
reduces~to\begin{equation}\left\langle\chi\left|\frac{\bm{p}^2}
{\sqrt{\bm{p}^2+m_1^2}}+\frac{\bm{p}^2}{\sqrt{\bm{p}^2+m_2^2}}
\right|\chi\right\rangle=\left\langle\chi\left|\,\bm{x}\cdot
\frac{\partial\,V}{\partial\bm{x}}(\bm{x})\,\right|\chi\right
\rangle.\end{equation}

\subsection{Trivial Upper Bounds: Nonrelativistic Limit of
Spinless-Salpeter Hamiltonians}\label{NRL}The (undeniable)
concavity of the square-root operator of the proper relativistic
expression~for the kinetic energy regarded as a function of
$\bm{p}^2$ implies that its nonrelativistic limit is tangent to
its relativistic precursor at the point of tangency $\bm{p}^2=0$.
Thus, the nonrelativistic Hamiltonian\begin{equation}H_{\rm NR}=
m_1+m_2+\frac{\bm{p}^2}{2\,\hat m}+V(\bm{x})\ ,\qquad\hat m\equiv
\frac{m_1\,m_2}{m_1+m_2}\ ,\label{Eq:HNR}\end{equation}forms an
upper bound to its originator~(\ref{H}), with expectable impact on
their sets of eigenvalues:\begin{equation}H\le H_{\rm NR}\qquad
\Longrightarrow\qquad\ E_k\le E_{{\rm NR},k}\ ,\qquad k\in{\mathbb
N}_0\ .\label{Eq:NRB}\end{equation}With respect to the total
number $N$ of bound states of the spinless Salpeter equation, the
above observation implies that this number can never be less than
its nonrelativistic counterpart \cite{WL:NR}.

\section{Informative Application to Two Spinless Semirelativistic
Problems}\label{SSP}In the past, the \emph{effortlessness\/} of
applying all the findings collected in Sect.~\ref{SSH} to
semirelativistic bound-state problems has been demonstrated
\cite{WL:WS,WL:Q@W,WL:H,WL:Y,WL:K,WL:Hm,WL:R} for a number of
frequently~employed, rather popular interaction potentials.
Therefore, let's take a brief look at two of these analyses, where
simplicity motivated us to confine ourselves to spherically
symmetric central potentials\begin{equation}V(\bm{x})=V(r)\
,\qquad r\equiv|\bm{x}|\ .\label{cp}\end{equation}

\subsection{Short-Range Singular Example: Spinless Relativistic
Hulth\'en Problem \cite{WL:Q@W,WL:H}}\label{SPo}
\begin{figure}[ht]\centering\includegraphics[scale=1.7076,clip]
{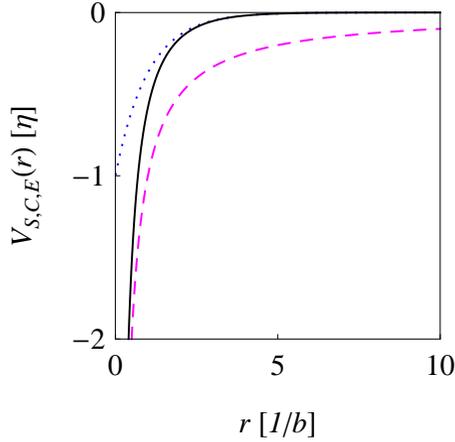}\\[1.52405ex]\caption{Hulth\'en potential $V_{\rm
S}(r)$ (solid black line), Eq.~(\ref{SP}), and asymptotes
(\ref{SCE}): negative exponential potential $V_{\rm E}(r)$ (dotted
blue line) and Coulomb potential $V_{\rm C}(r)$ (dashed magenta
line), for strength $\eta=1$.}\label{SPA}\end{figure}

The Hulth\'en
potential $V_{\rm S}(r)$, originally introduced in nuclear physics
but since then in physics widely used, is defined by two
parameters, the coupling strength $\eta$ and an exponential
range~$b$:\begin{equation}V_{\rm S}(r)\equiv-\frac{\eta}
{\exp(b\,r)-1}\ ,\qquad b>0\ ,\qquad\eta\ge0\ .\label{SP}
\end{equation}As shown in Fig.~\ref{SPA}, the potential $V_{\rm
S}(r)$ approaches, for short distances, from above a~Coulomb
potential $V_{\rm C}(r)$ with strength $\eta/b$ (i.e., exhibits
Coulombic behaviour) and, for large distances, from below a
negative exponential potential $V_{\rm E}(r)$, or increases
exponentially damped to zero:
\begin{equation}V_{\rm S}(r)\xrightarrow[r\to0]{}V_{\rm
C}(r)\equiv-\frac{\eta}{b\,r}\ ,\qquad V_{\rm S}(r)\xrightarrow
[r\to\infty]{}V_{\rm E}(r)\equiv-\eta\exp(-b\,r)\ .\label{SCE}
\end{equation}Conversely, the Hulth\'en potential (\ref{SP}) is
bounded from below by the Coulomb potential (\ref{CP})~iff
\begin{equation}\alpha\ge\frac{\eta}{b}\ .\label{CL}\end{equation}

Consequently, the Hulth\'en potential (\ref{SP}) inevitably
exhibits a Coulomb-like singularity at spatial origin $r=0$.
Nevertheless, according to Subsect.~\ref{BB} the assigned
Hamiltonian (\ref{H})~and bound-state spectrum are bounded from
below if the two involved potential parameters satisfy
\begin{equation}\frac{\eta}{b}<\frac{4}{\pi}\ .\label{SBB}
\end{equation}

For the \emph{nonrelativistic\/} limit (\ref{Eq:HNR}) of the
Hamiltonian (\ref{H}) with Hulth\'en potential (\ref{SP}) and all
states of orbital angular momentum $\ell=0$, the eigenvalues
$E_{\rm NR}$ can be found analytically~\cite{SF}:\begin{equation}
E_{{\rm NR},n}=m_1+m_2-\frac{\left(2\,\hat m\,\eta-n^2\,b^2
\right)^2}{8\,\hat m\,n^2\,b^2}\ ,\qquad\hat m\equiv
\frac{m_1\,m_2}{m_1+m_2}\ ,\qquad n\in{\mathbb N}\ ,\label{F}
\end{equation}where the \emph{radial\/} quantum number $n$ is
constrained by the two potential characteristics $\eta$ and~$b$,
\begin{equation}n\le\frac{\sqrt{2\,\hat m\,\eta}}{b}\ .\label{n}
\end{equation}In the spirit of Subsect.~\ref{NRL}, the expressions
(\ref{F}) may serve to provide, at least, a vague idea of the
actual location of the discrete eigenstates of any spinless
semirelativistic Hulth\'en~problem.

\subsection{Category of Spinless Semirelativistic
Generalized-Hellmann Problems \cite{WL:Hm,WL:R}}Generalized
Hellmann potentials \cite{WL:Hm,WL:R} establish a class of central
interaction potentials that encompasses the shape introduced, some
time ago \cite{H1,H2}, for applications in atomic-physics theory.
Each member of this set is defined to be a linear combination of
an attractive Coulomb contribution $V_{\rm C}(r)$ and an
optionally \emph{either attractive or repulsive\/} Yukawa
contribution $V_{\rm Y}(r)$,\begin{equation}V_{\rm H}(r)\equiv
V_{\rm C}(r)+V_{\rm Y}(r)=-\frac{\kappa}{r}-\upsilon\,
\frac{\exp(-b\,r)}{r}\label{HP}\ ,\end{equation}Coulomb coupling
$\kappa$, Yukawa coupling $\upsilon$, and exponential-range
parameter $b$ hence satisfying\begin{equation}\kappa\ge0\ ,\qquad
\upsilon\gtreqqless0\ ,\qquad b>0\ .\label{PC}\end{equation}(Note
that, in the notation (\ref{HP}), the interaction potential
originally proposed by Hellmann \cite{H1} corresponds to assuming
for the Yukawa coupling parameter $\upsilon$ a strictly negative
value $\upsilon<0$.) At large distances $r$, all generalized
Hellmann potentials approach their Coulomb~component,
\begin{equation}V_{\rm H}(r)\xrightarrow [r\to\infty]{}V_{\rm
C}(r)\ .\label{0}\end{equation}This guarantees that any discrete
eigenvalue $E_k$ of $H$ is nonpositive, thus bounded from~above:
\begin{equation}E_k\le0\qquad\forall\quad k\in{\mathbb N}_0\
.\label{EH}\end{equation}Due to the asymptotic approach (\ref{0})
to Coulomb-type behaviour, every generalized Hellmann potential
(\ref{HP}) fails to meet all requirements imposed in
Subsect.~\ref{nde}, particularly, because \cite{WL:Hm}
\begin{equation}V_{\rm H}(r)\notin L^{3/2}({\mathbb R}^3)\cap
L^3({\mathbb R}^3)\ ,\label{nin}\end{equation}whence no finite
upper bound (\ref{D}) on the total number of its discrete bound
states can be~found.

\emph{Nonsingular\/} generalized Hellmann potentials, in the
classification of Table~\ref{GHP} identified by
$\kappa+\upsilon\le0$, are bounded from below, as likewise
(because of $\sqrt{\bm{p}^2+m^2}\ge0$) their Hamiltonians:
\begin{alignat}{3}H\ge V_{\rm H}(r)&\ge\min_{0\le r<\infty}V_{\rm
H}(r)>-\infty&\qquad\mbox{for}\qquad&\upsilon<-\kappa\ ,&\label{r}
\\[.3ex]H\ge V_{\rm H}(r)&\ge V_{\rm H}(0)=\upsilon\,b&\qquad
\mbox{for}\qquad&\upsilon=-\kappa\ .&\label{f}\end{alignat}

\emph{Singular\/} generalized Hellmann potentials, in their
systematics of Table~\ref{GHP} characterized by
$\kappa+\upsilon>0$, develop \emph{negative singularities\/} at
$r=0$ bounded from below by a Coulomb~potential:\begin{equation}
V_{\rm H}(r)\ge-\frac{\alpha}{r}\qquad\mbox{with}\qquad
\left\{\begin{array}{ll}\alpha=\kappa+\upsilon&\qquad\mbox{for}
\qquad\upsilon>0\ ,\\[.5ex]\alpha=\kappa&\qquad\mbox{for}\qquad
\upsilon\le0\ .\end{array}\right.\label{CLB}\end{equation}As
consequence of this, \emph{mutatis mutandis\/} the relevant
insights collected in Subsect.~\ref{BB} apply. Most importantly,
spinless relativistic generalized-Hellmann problems are well
defined \cite{WL:Hm}~if\begin{equation}\kappa+\upsilon\le
\frac{4}{\pi}\ .\end{equation}

\begin{table}[hbt]\centering\caption{Characterization of the seven
types of generalized Hellmann potentials $V_{\rm H}(r)$
distinguished by comparing the Yukawa coupling $\upsilon$ with the
(by assumption nonnegative) Coulomb coupling $\kappa\gneqq0$
\cite{WL:Hm}.}\label{GHP}\vspace{1ex}\begin{tabular}{lllcc}\toprule
Boundedness&Characteristic&Behaviour near&\multicolumn{1}{l}{Sign
of sum}&\multicolumn{1}{l}{Relation between}\\from below&of
potential&the origin $r=0$&\multicolumn{1}{l}{of couplings}&
\multicolumn{1}{l}{couplings $\upsilon$ and $\kappa$}\\\midrule
unbounded&``singular''&$V_{\rm H}(r)\xrightarrow[r\to0]{}-\infty$
&$\kappa+\upsilon>0$&$\upsilon>\kappa$\\&&&&$\upsilon=\kappa$\\
&&&&$0<\upsilon<\kappa$\\&pure Coulomb&&&$\upsilon=0$\\&&&&
$-\kappa<\upsilon<0$\\\midrule bounded&finite at origin&$V_{\rm
H}(r)\xrightarrow[r\to0]{}\upsilon\,b$& $\kappa+\upsilon=0$&
$\upsilon=-\kappa$\\&repulsive core&$V_{\rm H}(r)
\xrightarrow[r\to0]{}+\infty$&$\kappa+\upsilon<0$&
$\upsilon<-\kappa$\\\bottomrule\end{tabular}\end{table}

\begin{figure}[ht]\centering{\begin{tabular}{cc}
\includegraphics[scale=1.7076,clip]{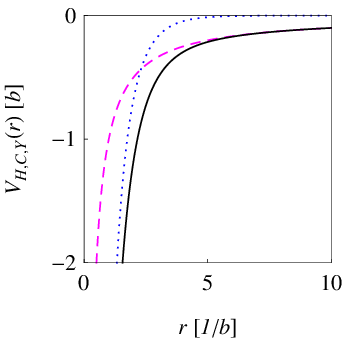}\hspace{1ex}&\hspace{1ex}
\includegraphics[scale=1.7076,clip]{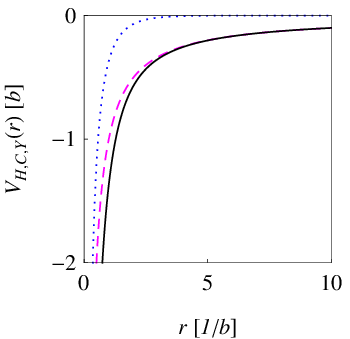}\\[2ex](a)&(b)\\[4.3849ex]
\includegraphics[scale=1.7076,clip]{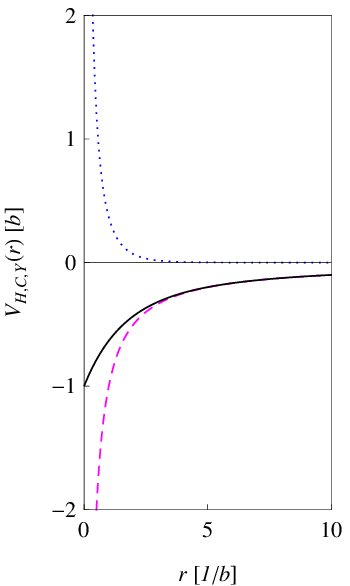}\hspace{1ex}&\hspace{1ex}
\includegraphics[scale=1.7076,clip]{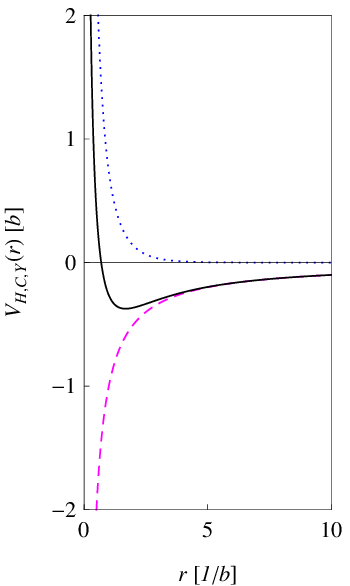}\\[2ex](c)&(d)\end{tabular}}
\\[1.52405ex]\caption{Generic representatives (solid black lines)
of four of the only seven distinguishable categories in our
systematic classification (Table~\ref{GHP}) of generalized
Hellmann potentials $V_{\rm H}(r)$ defined in Eq.~(\ref{HP}) by
linear combinations of Coulombic (dashed magenta lines) and Yukawa
(dotted blue lines) contributions, choosing $\kappa=1$ and (a)
$\upsilon=10$, (b) $\upsilon=1$, (c) $\upsilon=-1$, or (d)
$\upsilon=-2$ for the involved coupling strengths.}\label{Pos}
\end{figure}

Merely for illustrative purposes, Fig.~\ref{Pos} depicts a typical
representative of four of our seven classes of generalized
Hellmann potentials (\ref{HP}) identified in Table~\ref{GHP}.
There is no need to show the trivial case of the pure Coulomb
potential arising for vanishing Yukawa coupling~strength.

In order to exemplify the calculation of upper bounds on energies
along the lines sketched in Subsect.~\ref{U}, Table~\ref{B}
provides, for few low-lying eigenstates of three
generalized-Hellmann problems and a fixed choice of the numerical
values of the parameters $\mu$ and $\beta$ in the basis (\ref{LB})
of some $d$-dimensional subspace, upper bounds on the associated
binding energies,~defined~by\begin{equation}B_k\equiv E_k-2\,m\
,\qquad k\in{\mathbb N}_0\ .\label{b}\end{equation}The upper
bounds in Table~\ref{B} can straightforwardly be optimized
variationally, in two respects: definitely by increasing the
dimension $d$ of the adopted subspace, and potentially by trying
out different values for (i.e., varying) the parameters $\mu$
and/or $\beta$ defining the subspace's basis~(\ref{LB}).

\clearpage

\begin{table}[hb]\centering\caption{Bounds on binding energies,
for states of radial excitation $n_{\rm r}$ and orbital angular
momentum $\ell$, for exemplary spinless relativistic
generalized-Hellmann problems \cite{WL:Hm,WL:R} ($b=\mu=m$,
$\beta=1$,
$d\gtrapprox29$).}\label{B}\begin{tabular}{cclll}\toprule&&
\multicolumn{3}{c}{Upper bounds on $B_{n_{\rm r}\ell}\;[m]$}\\
\cline{3-5}\\[-2ex]\multicolumn{2}{c}{Bound state}&\multicolumn{1}
{c}{$\kappa=\upsilon=\frac{1}{2}$}&\multicolumn{1}{c}{$\kappa=1$,
$\upsilon=-1$}&\multicolumn{1}{c}{$\kappa=1$, $\upsilon=-2$}\\
\cline{1-2}\\[-2ex]$n_{\rm r}$&$\ell$&\multicolumn{1}{c}{[case
Fig.~\ref{Pos}(b)]}&\multicolumn{1}{c}{[case Fig.~\ref{Pos}(c)]}&
\multicolumn{1}{c}{[case Fig.~\ref{Pos}(d)]}\\\midrule
0&0&$\quad\!\!-0.11673$&$\quad-0.17951$&$\quad-0.14410$\\
0&1&$\quad\!\!-0.01579$&$\quad-0.06294$&$\quad-0.06157$\\
0&2&$\quad\!\!-0.00616$&$\quad-0.02813$&$\quad-0.02812$\\[.8ex]
1&0&$\quad\!\!-0.02107$&$\quad-0.05464$&$\quad-0.04786$\\
1&1&$\quad\!\!-0.00509$&$\quad-0.02810$&$\quad-0.02762$\\[.8ex]
2&0&$\quad\!\!-0.00688$&$\quad-0.02566$&$\quad-0.02338$\\
\midrule\multicolumn{2}{l}{Lower bound on $B_{n_{\rm
r}\ell}\;[m]$}&$\quad\!\!-0.58578\dots$&$\quad-1$&
$\quad-0.37336\dots$\\\bottomrule\end{tabular}\end{table}

\nocite{*}
\bibliography{PESA}
\bibliographystyle{woc}
\end{document}